# Ultrathin, polarization-independent, and focus-tunable liquid crystal diffractive lens for augmented reality

Mareddi Bharath Kumar[1], Daekyung Kang[1], Jihoon Jung[2], Hongsik Park[2], Joonku Hahn[2], Muhan Choi[2], Jin-Hyuk Bae[2], Hyunmin Kim[3], and Jonghoo Park[1*]

[1]Department of Electrical Engineering, Kyungpook National University, Daegu, Republic of Korea

[2]School of Electronics Engineering, Kyungpook National University, Daegu, Republic of Korea

[3]Companion Diagnostics & Medical Technology Research Group, DGIST, Daegu 42988, Republic of Korea.

*Correspondence to: jonghoopark@knu.ac.kr

**Abstract**

Despite the recent advances in augmented reality (AR), which has shown the potential to significantly impact on our daily lives by offering a new way to manipulate and interact with virtual information, minimizing visual discomfort due to the vergence–accommodation conflict remains a challenge. Emerging AR technologies often exploit focus-tunable optics to address this problem. Although they demonstrated improved depth perception by enabling proper focus cues, a bulky form factor of focus–tunable optics prevents their use in the form of a pair of eyeglasses. Herein, we describe an ultrathin, focus-tunable liquid crystal (LC) diffractive lens with a large aperture, a low weight, and a low operating voltage. In addition, we show that the polarization dependence of the lens, which is an inherent optical property of LC lenses, can be eliminated using birefringent thin films as substrates and by aligning the optical axes of the birefringent substrates and LC at a specific angle. The polarization independence eliminates the need for a polarizer, thus further reducing the form factor of the optical system. Next, we demonstrate a prototype of AR glasses with addressable focal planes using the ultrathin lens. The prototype AR glasses can adjust the accommodation distance of the virtual image, mitigating the vergence–accommodation conflict without substantially compromising the form factor or image quality. This research on ultrathin lens technology shows promising potential for developing compact optical displays in various applications.

Augmented reality (AR) is a technology that overlays computer–generated virtual information on a real-world environment [1,2]. It allows users to manipulate and interact with virtual objects in the context of the real world around them. The interface between an AR system and a user has evolved to the form of a pair of eyeglasses, offering users a more immersive experience in various applications including entertainment, education, training, and marketing. However, conventional AR systems have been known to suffer from the vergence–accommodation conflict for the past decades [3,4], which causes visual discomfort such as visual fatigue [5] and depth perception error [4,6-9]. Vergence is a binocular eye movement wherein the eyes are rotated in opposite directions to direct the visual axis of each eye on the same object. Accommodation refers to the adjustment of the focal power (i.e., shape) of a crystalline lens to obtain a clear image at different depths. In a natural environment, vergence and accommodation are strongly coupled. Therefore, the vergence distance, which is the distance between the eyes and a point where two visual axes intersect, coincides with the accommodation distance (i.e., the focal distance). In conventional AR systems, a virtual image is formed at different stereoscopic distances by displaying two slightly different images separately on each eye using see-through near-eye displays. This artificially induced binocular disparity leads to vergence, yielding a different vergence distance. However, the accommodation distance is fixed at the optical distance of the displays. Therefore, a user may not be able to see the real and virtual objects in focus simultaneously. This limitation is further exacerbated when augmenting a relatively close real object with a virtual image or information, such as in surgical training [10].

Among the various AR technologies that have been explored to address this problem [11-25], an interesting approach is the use of variable focal plane near-eye displays [10,26-32]. In these systems, the focus-tunable optics, mostly a liquid lens, is placed in the optical path of the virtual image. The optical power of the focus-tunable lens is adjusted to match the focal distance of the virtual image with the vergence distance. Such systems exhibit improved depth perception [9,10,26-31,33] and significantly reduce the time required to fuse a pair of stereoscopic images [34] in a human visual system. However, their utility is compromised by the bulky form factor and high operating voltage of the tunable lenses. Thus, these systems are far from being suitable for practical use in the form of a pair of eyeglasses. An ideal focus-tunable lens for compact wearable AR glasses should be thin and lightweight while having a broad focus-tunable range, a large aperture, and a low operating voltage.

Herein, we report an ultrathin (~266 μm), polarization-independent, and focus-tunable liquid crystal (LC) diffractive lens with a large area (a diameter of 20 mm), a low weight (1.5 g), and a low operating voltage (< 2.1 V). The lens was implemented using a nematic LC as an active layer and birefringent thin films as substrates. We show that a proper alignment of the optical axes of the birefringent substrates, which meet quarter-wave plate conditions, and LC helps to eliminate the polarization dependence of the LC diffractive lens. In addition, we demonstrate the RGB imaging characteristics of the lens for 10 discrete, switchable optical powers ranging from −3 D to +3 D. Next, we demonstrate a monocular prototype of a near-eye see-through AR system implemented by integrating our focus-tunable lens with commercial AR glasses. The prototype AR glasses can adjust the accommodation distance of the virtual image, providing improved depth perception without substantially compromising the form factor or image quality.

Figure 1a shows a schematic of the structure of the lens. The homogeneously aligned nematic LC is sandwiched between two birefringent substrates. The fabrication of the bottom substrate started from the indium tin oxide (ITO)-coated 127 μm-thick polyethylene terephthalate (PET) film. (See supplementary information section I for a detailed fabrication procedure). The 130 nm-thick ITO was patterned to form Fresnel zone pattern electrodes by photolithography. The Fresnel zone pattern electrodes consist of 46 Fresnel zones, with each Fresnel zone divided into 12 subzones with a 1 μm gap. The outer radii of each Fresnel zone [35] and subzone [36] were determined corresponding to an optical power of 0.5 D and a phase difference of 2π at each Fresnel zone boundary for a wavelength of 543 nm. As shown in Figure 1b, the subzones with the same indices in all the Fresnel zones are electrically connected by an aluminum interconnect line (black lines) through via holes (red dots) on an SU-8 insulation layer. The Al lines are then extended to the contact pads. The top substrate consists of 130 nm-thick ITO coated on a PET as the ground electrode and poly(vinyl alcohol) (PVA) as an LC alignment layer. The PVA layers on the top and bottom substrates were rubbed in an antiparallel direction using a velvet cloth. The two substrates were bonded using 10 μm-thick adhesive spacers placed on the rim of one substrate. A commercial nematic LC (E7) was filled between the two substrates by capillary action. The thickness of the LC was uniformly maintained over the lens area using bead spacers (diameter: 10 μm) sprayed between the two substrates. Figure S1 shows the fabricated lens. As both the PET and LC are made of a birefringent material, the optical properties of the lens strongly depend on their thickness and optical axis alignment. To eliminate the polarization dependence of the lens, the extraordinary axis of the LC was aligned with those of the upper and lower PET films at 45°, while the optical axes of the upper and lower PET films were made perpendicular to each other as shown in Figure 1a. In addition, the thickness of the PET films should meet the quarter-wave plate conditions.

To demonstrate the polarization independence of the lens, the voltage-dependent optical transmission of the homogeneously aligned LC sandwiched between the PET substrates was measured. For a direct comparison, the same measurement was performed for a conventional structure comprising a homogeneously aligned LC sandwiched between two glass substrates (glass/LC/glass). The measurements were performed in crossed polarizers for different angles ($\varphi$) between the extraordinary axis of the LC and the optical axis of the polarizer while varying the voltage applied to the LC (see supplementary information section II). Figure 2a shows the maximum (red line) and minimum (blue dash line) optical transmission for the conventional structure, i.e., glass/LC/glass. The voltage-dependent optical transmission, which is translated into phase retardation, strongly depends on the angle between the optical axes of the LC and the polarizer. The electro-optic effects only arise for certain polarization states of the incident light, i.e., at angles of odd number multiples of π/4. For the angles of $\varphi$ that are even number multiples of π/4, the optical transmission remains zero, regardless of the voltage applied to the LC. The polarization dependence problem in the conventional structure can be solved either by stacking two lenses with orthogonal extraordinary optical axes or by placing a polarizer in front of the lens [37]. However, both approaches decrease the optical transmittance and increase the total thickness of the lens substantially. Figure 2b shows the numerical prediction, calculated using the Jones matrix, of the voltage-dependent optical transmission of the glass/LC/glass structure for different angles ($\varphi$). The prediction is in good agreement with the measurement results shown in Figure 2a. Figure 2c shows the maximum (red line) and minimum (blue dash line) optical transmission values of the LC sandwiched between the PET substrates with the optical axes arranged as shown in Figure 1a, i.e., (PET(0°)/LC(45°)/PET(90°)). Unlike that in the conventional structure, the electro-optic effect is induced for all angles of $\varphi$, indicating polarization independence of the structure. As shown in Figure S3a, the optical transmission oscillates between the minimum and maximum values as a function of the voltage applied to the LC. Figure S3b shows the phase retardation as a function of the voltage extracted from Figure S3a. The thickness and birefringence of the PET are 127 μm and 0.02215, respectively. This yields a phase retardation of 65° for a wavelength of 543 nm. An ideal polarization-independent lens should have a constant maximum and minimum optical transmission for all angles of $\varphi$. This can be achieved using a PET with a thickness of 128.5 μm, which meets the quarter-wave plate conditions. Figure 2d shows the numerical predictions of the voltage-dependent optical transmission for the arrangement PET(0°)/LC(45°)/PET(90°), calculated using the following Jones matrix (see supplementary information section III):

$$\begin{bmatrix} \cos\frac{\Delta_2(V)}{2}+i\sin\frac{\Delta_2(V)}{2}\sin2\gamma\cos\Delta_1 & 2\sin\frac{\Delta_1}{2}\cos\frac{\Delta_1}{2}\sin\frac{\Delta_2(V)}{2}-i\sin\frac{\Delta_2(V)}{2}\cos2\gamma\cos\Delta_1 \\ -2\sin\frac{\Delta_1}{2}\cos\frac{\Delta_1}{2}\sin\frac{\Delta_2(V)}{2}-i\sin\frac{\Delta_2(V)}{2}\cos2\gamma\cos\Delta_1 & \cos\frac{\Delta_2(V)}{2}-i\sin\frac{\Delta_2(V)}{2}\sin2\gamma\cos\Delta_1 \end{bmatrix},$$

where $i$ is the imaginary unit, $\Delta_1=2\pi\cdot\delta n\cdot d_{PET}/\lambda$ is the phase retardation of the PET, $\delta n$ and $d_{PET}$ are the birefringence and thickness of the PET, respectively, $\Delta_2=2\pi\cdot\delta n(V)\cdot d/\lambda$ is the phase retardation of the LC, $\delta n(V)=n_{eff}(V)-n_0$, $1/n_{eff}^2(V) = \cos^2\theta(V)/n_e^2 + \sin^2\theta(V)/n_o^2$, $n_e$ and $n_o$ are the extraordinary and ordinary refractive indices of the LC, respectively, and $\gamma$ is the angle between the ordinary axis of the PET and the optical axis of the polarizer. θ(V) is the tilt angle between the plane perpendicular to the propagating direction of the ray and the extraordinary axis of the LC, V is the voltage across the LC, d is the thickness of the LC, and λ is the wavelength of the incident light. Two values of phase retardation of the PET ($\Delta_1$) were used in the calculation: the measured phase retardation of the PET and an odd number multiple of π/2. The calculation results for a phase retardation of 65° (red line) agrees well with the experimental results shown in Figure 2c. With $\Delta_1$, which meets either one or three quarter-wave plate conditions, the polarization dependence was eliminated (green dash line), as the above matrix becomes independent of φ (see supplementary information section III). Achieving polarization independence eliminates the need for a polarizer, thus further reducing the form factor of the optical system.

Although the lens was designed to provide an optical power of 0.5 D with 12 discrete phase levels per a phase difference of 2π, it can be electrically reconfigured to provide optical powers of ±1.0 D with six phase levels, ±1.5 D with four phase levels, ±2.0 D with three phase levels, and ±3.0 D with two phase levels, as shown in Figure 3a and Figure S5 (for ±3.0 D with two phase levels). To demonstrate the characteristics of the lens, i.e., whether the optical power can be changed electrically, the images of a target located at different distances were captured using a model eye comprising an achromatic refractive lens (with a focal length of 19.0 mm) and a charge coupled device (CCD). The LC lens was placed in front of the model eye. The images were taken under white light illumination without a wavelength filter or a polarizer. Figure 3b shows the images of the USAF target printed in RGB colors on a photo paper, captured when the LC lens was turned off (*i*) and on (*ii-v*) with different optical powers. The target, which was placed at 48, 62, 93, and 193 cm away from the model eye, was brought into focus when the lens was on for optical powers of 2.0, 1.5, 1.0, and 0.5 D, respectively. To obtain images for a negative LC lens power, a meniscus lens with an opposite power was placed in front of the LC lens. The target located at a fixed distance of 60 cm away from the model eye is first focused using the model eye and then defocused by placing a positive power meniscus lens in front of the LC lens, as shown in the first image (*i*) in Figure 3c. Finally, the target is brought into focus again by compensating for the optical power of the meniscus lens using the negative power LC lens, as shown in Figure 3c (*ii-v*). Figure 3d shows the modulation transfer function (MTF) of the entire optical system comprising the LC lens, a refractive lens, and a CCD with a resolution of 1024 × 768 pixels. The MTFs for different optical powers were determined by analyzing the images of slanted edges in the ISO 12233 resolution chart obtained under each LC lens power. The MTF 50 slightly decreases with the increase in the optical power, as shown in the inset of Figure 3d. This is because the higher the optical power, the lower the diffraction efficiency. The MTF decreases compared to that of the model eye because the phase retardation of the PET substrate does not perfectly meet the one or three quarter-wave plate condition; nevertheless, this can be improved by engineering the thickness and birefringence of the PET, as shown in Figure 2d.

To implement compact near-eye see-through AR glasses capable of controlling the depth of a virtual image, the LC lens is placed at the eye side of the commercial AR glasses, as shown in Figure 4a. It is noteworthy that only one ultrathin, lightweight LC lens was incorporated in the commercial AR glasses. Ideally, an LC lens should be placed in the optical path of the virtual image so that the AR glasses maintain the accommodation distance of the real object while correcting that of the virtual image using the LC lens. However, in this prototype, we deliberately utilize the zeroth and first diffraction orders of the LC lens to produce two foci: one for a nearer object and one for a farther object. When the lens is turned on, the zeroth-order energy of the incident light is directed at the focal point of the refractive lens, whereas the first-order energy is directed at the effective focal point formed by the combination of the refractive and LC lenses. The first diffraction order efficiencies of the LC lens corresponding to optical powers of −2 D (three phase levels), −1.5 D (four phase levels), and −1 D (six phase levels) are 68.4, 81.1, and 91.2%, respectively, in theory [38]. The remaining incident light is distributed between the zeroth and higher orders.

To simulate an AR training task performed within arm's reach, for example in surgical training, a real object (a teddy bear) was placed 45 cm away from the AR glasses. The virtual image (a heart) was displayed by the commercial near-eye see-through AR glasses at a focal distance of 310 cm. As shown in Figure 4b, the real object is seen in focus when focusing the model eye on the real object with the LC lens turned off, whereas the virtual image is out of focus because of the mismatch between their accommodation distances. When the lens is turned on with an optical power of −2 D, the focal distance of the virtual image is decreased to 45 cm, bringing both the real and virtual objects in focus. Figure 4c shows the depth control of the virtual image at other positions for LC lens powers of −1.5 and −1 D, respectively. Three real targets were placed at 50 (K in red), 69 (N in green), and 97 cm (U in blue) away from the AR glasses as references to the depth of the virtual image. When focusing the model eye at 69 and 97 cm, respectively, both the real and virtual toruses were seen in focus for LC lens power of −1.5 and −1 D, respectively. The insets show the images captured when the LC lens was turned off; the virtual torus is out of focus while the real target (K, N, U) is in focus.

Despite the recent advancements in resolving the vergence–accommodation conflict in AR systems using focus-tunable optics, reducing their form factor remains a key challenge for developing consumer-grade AR platforms. We report a substantial progress in the development of focus tunable lenses that significantly reduce the form factor of AR systems by not only reducing the thickness of the lens but also by eliminating the need for a polarizer and a wavelength filter. Our initial study suggests that this lens technology can be potentially applied to AR glasses for improved depth perception without substantially compromising the form factor or image quality. In the future, we plan to develop compact gaze-contingent and adaptive focusing AR glasses by integrating the lens with a gaze tracker. Moreover, time-multiplexed operating of the lens will be explored.

**Materials and methods**

Details regarding the fabrication procedure for the lens and measurement procedure for the voltage-dependent optical transmission are given in the supplementary information section. The commercial nematic LC (E7) and ITO-coated PET were purchased from INSTEC, Inc (USA) and Sigma Aldrich (USA), respectively. The birefringence of the PET was measured using a retardation film and material evaluation system (RETS-100, Otsuka Electronics Co., Ltd.) with a wavelength of 550 nm. The images were captured using the model eye comprising a achromatic refractive lens (with a focal length of 19.0 mm) and a CCD camera (DCU223C, Thorlabs) with a resolution of 1024 × 768 pixels. The 12 voltage waveforms used to drive the lens were generated using a PCI-6723 analog output module (NI). Each channel generates a 100 Hz bipolar pulse with zero DC bias. The MTF was calculated using Imatest Master 5.1.22 (Imatest LLC). Commercial AR glasses (BT-300, Epson) was used for the prototype. The image of the heart was purchased from 123rf.com.

**Acknowledgment**

This work was supported by Samsung Research Funding & Incubating Center of Samsung Electronics under project number SRFC-IT1301-51, SAMSUNG Research of Samsung Electronics Co., Ltd, and the National Research Foundation of Korea (NRF) grant funded by the Korea government (MSIP) (NRF-2017R1A4A1015565).

**Contributions**

J.P conceived the idea, designed the experiment, and directed the research. M.B.K, D.K, and J.J fabricated the lens. J.P, M.B.K, J.H, H.P, M.C, J-H.C, and H.K. performed the measurements. J.P and M.B.K wrote the paper.

**Conflict of interest**

The authors declare that they have no conflict of interest.

**Figure captions**

**Figure 1.** Schematics of the ultrathin, polarization-independent liquid-crystal diffractive lens: (a) Orientations of the extraordinary axes of the top and bottom PETs and the LC. Explored view of the LC lens highlighting the different components and thicknesses of the multilayer architecture. The first and second layers from the bottom illustrate a Fresnel zone-patterned ITO on a PET substrate and the aluminum interconnect lines patterned on an SU-8 layer, respectively. A nematic LC layer is sandwiched between the PVA layers rubbed in antiparallel directions. The unpatterned ITO on the top PET substrate is grounded. (b) Detailed view of the Fresnel zone-patterned electrodes, interconnect lines, and via holes. The subzones with the same indices are connected by an Al interconnect line through via holes.

**Figure 2.** Voltage-dependent optical transmission of the lens in crossed polarizers for different angles between the optical axes of the polarizer and the LC: (a) Measured and (b) calculated maximum (red solid line) and minimum (blue dash line) optical transmission for a conventional glass/LC/glass structure exhibiting polarization dependence. (c) Measured maximum (red solid line) and minimum (blue dash line) optical transmission for the PET(0°)/LC(45°)/PET(90°) structure, exhibiting electro-optic effects for all angles between the optical axes of the polarizer and the LC, indicating polarization independence. (d) Numerical prediction of the voltage-dependent optical transmission calculated using the Jones matrix for the PET(0°)/LC(45°)/PET(90°) structure with two different PET thicknesses. The prediction obtained using the measured PET thickness (127 μm, red solid line) agrees well with the measured result. When using the thickness of the PET that meets the quarter-wave plate condition (128.5 μm, green dash line), there is no polarization dependence.

**Figure 3.** RGB images with different positive and negative powers of the lens under white light illumination without a polarizer or a wavelength filter: (a) Electrically reconfigurable phase profile of the lens for different optical powers. (b) Images of USAF resolution with different positive powers of the lens. (*i, ii*) Images of the target placed at a reading distance (48 cm) when the lens is turned (*i*) off and (*ii*) on with +2.0 D. (*iii-vi*) Images of the target placed at 62, 93, and 193 cm, when the lens is turned on; the corresponding lens powers for focusing are +1.5, +1.0, and +0.5 D, respectively. (c) Images of USAF target with different negative powers of the lens. The target is fixed 60 cm away from the model eye, and a meniscus lens with opposite power is placed in front of the LC lens. (d) Modulation transfer function of the lens for different optical powers. The MTF 50 (inset) slightly decreases with the increase in the lens power.

**Figure 4.** Schematic of a compact monocular prototype of a see-through near-eye AR system and AR images obtained using the same: (a) Schematic of the compact near-eye see-through AR system. The ultrathin lens is placed between the model eye and the commercial AR glasses. (b) AR images obtained using the AR glasses. (left) AR image when the model eye focuses on the real object and the LC lens is turned off. The virtual image is out of focus because of the accommodation distance mismatch between the real and virtual objects placed at 45 and 310 cm, respectively. (right) When the LC lens is turned on with a power of −2 D, both the real and virtual objects are in focus. (c) AR images when the model eye focuses on the real object "N" placed at 69 cm with an LC lens power of −1.5 D. The inset shows the AR image when the LC lens is turned off. (d) AR image when the model eye focuses on the real

object "U" placed at 97 cm with a lens power of −1.0 D. The inset shows the AR image when the LC lens is turned off.

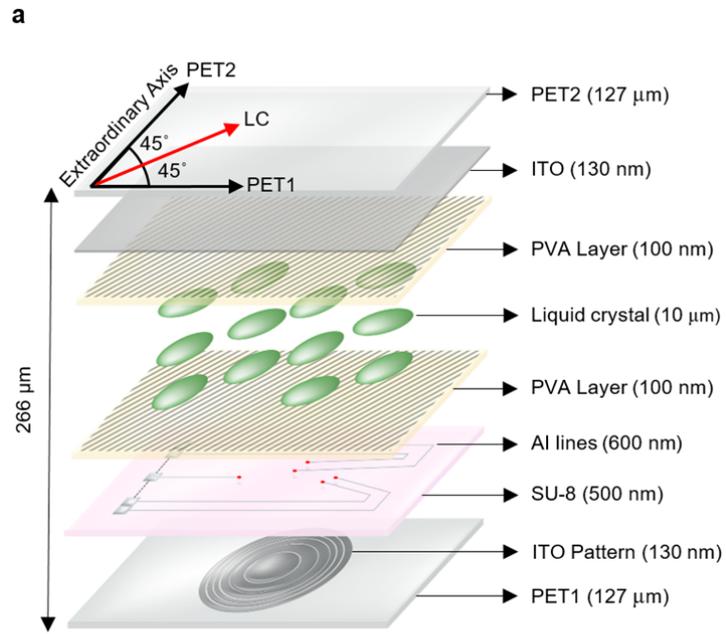

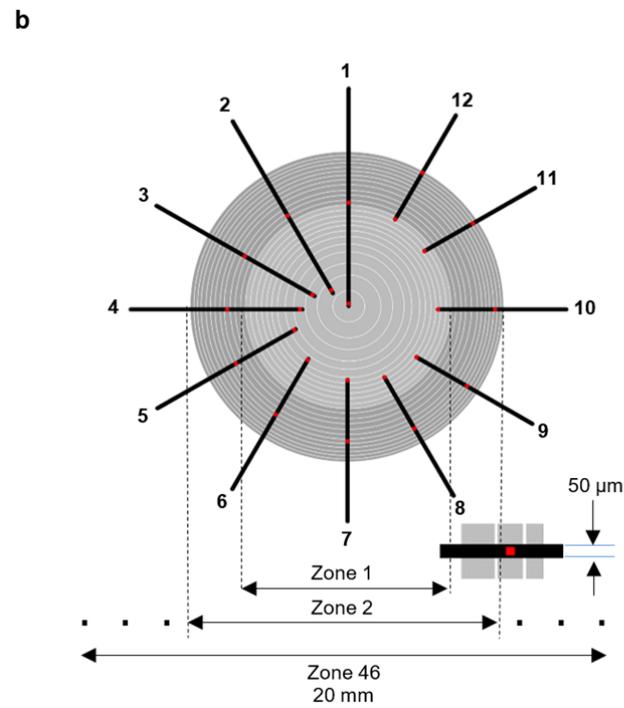

Figure 1. Kumar *et al*.

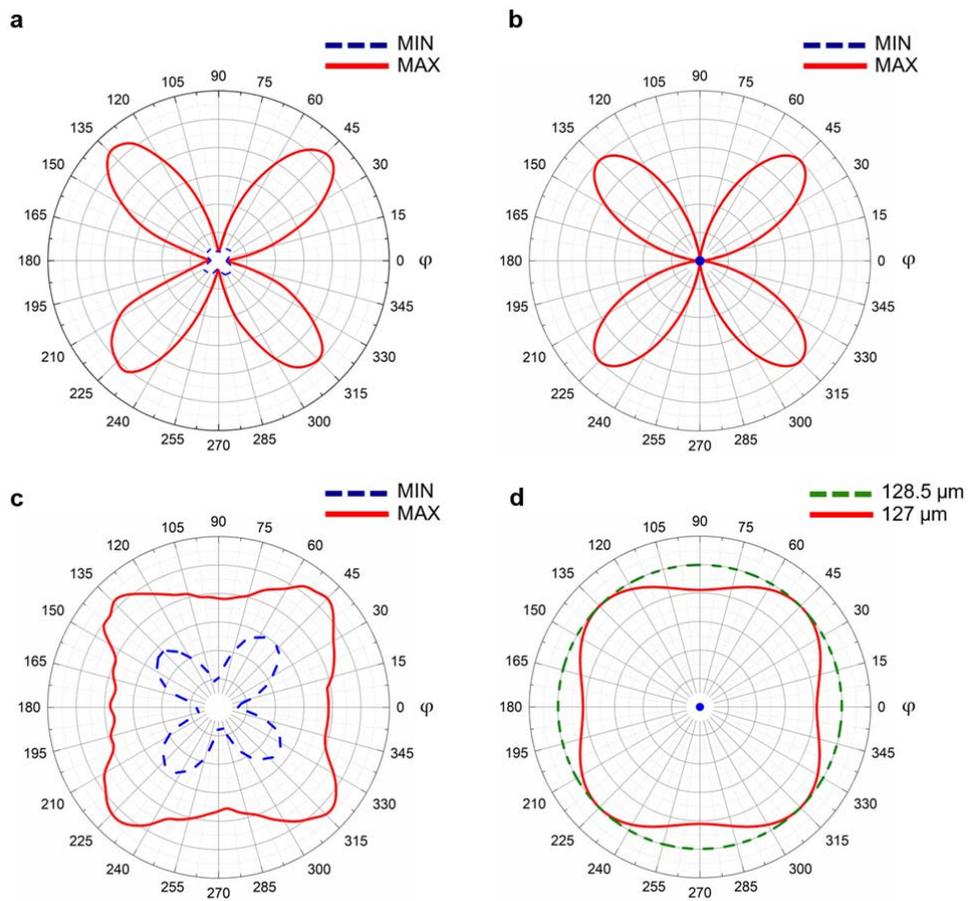

Figure 2. Kumar *et al*.

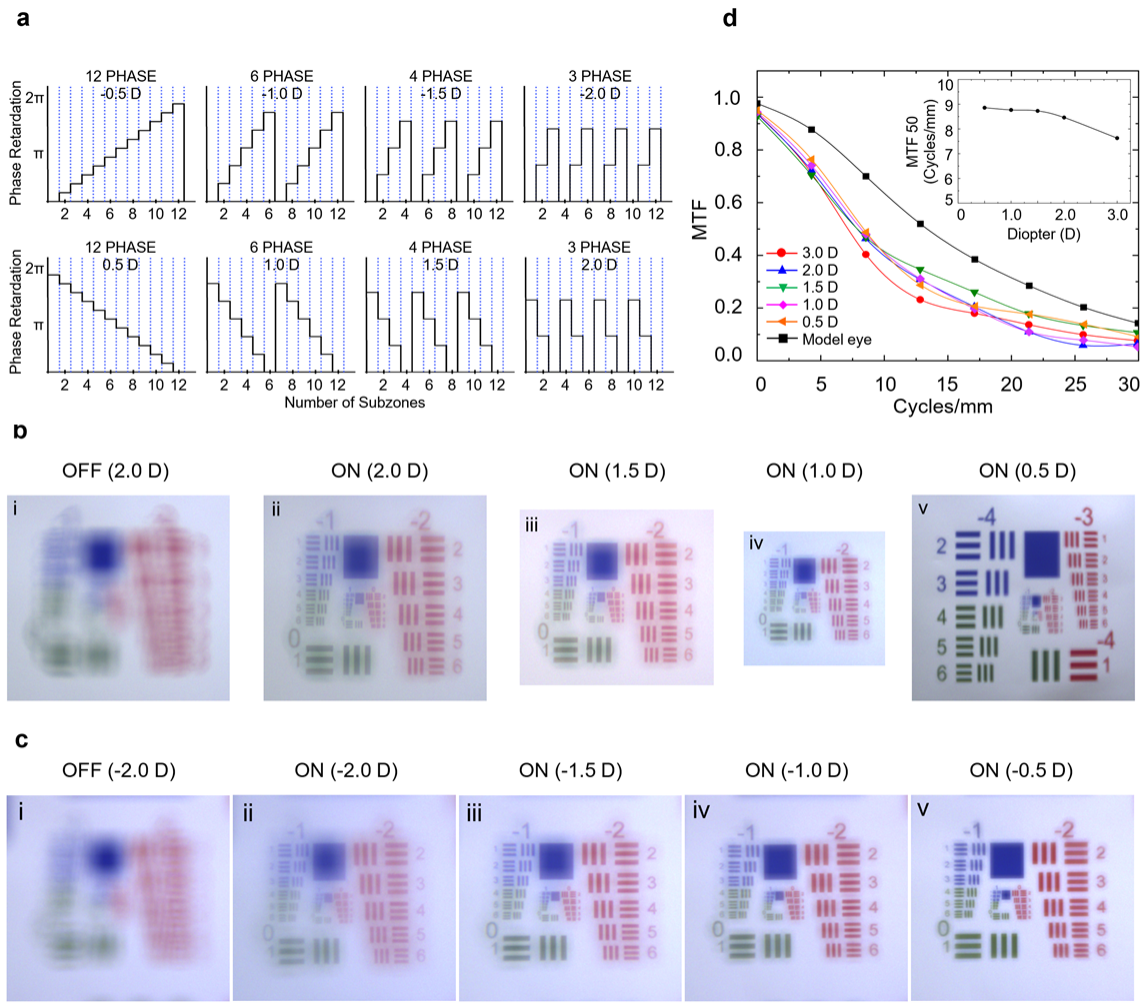

Figure 3. Kumar *et al*.

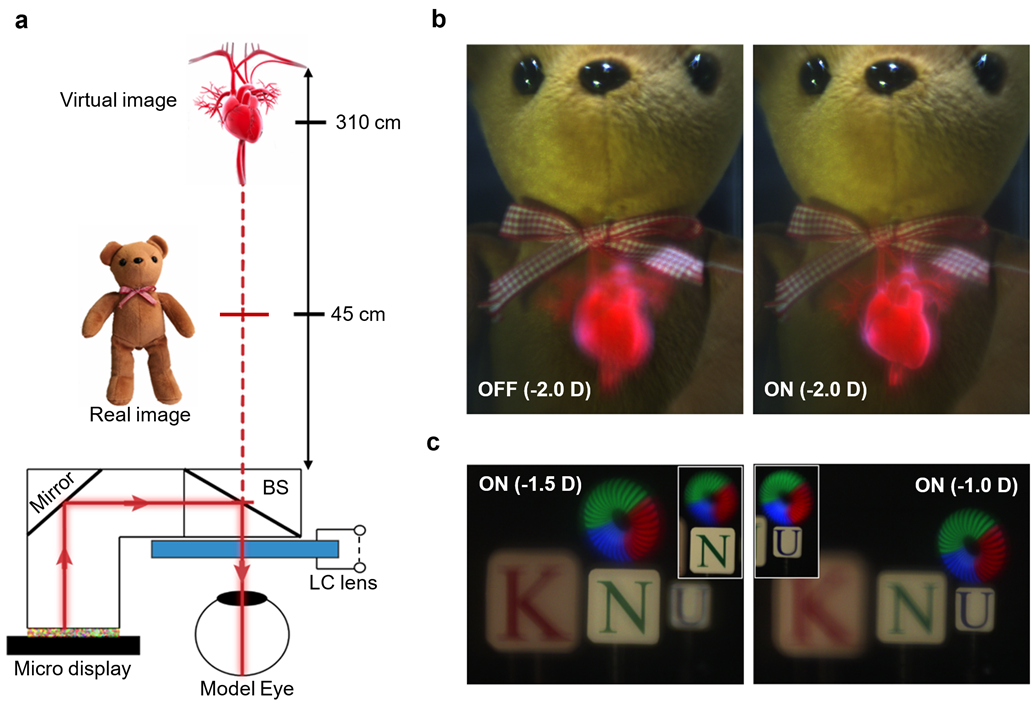

Figure 4. Kumar *et al*.

# Supplementary Information

## Ultrathin, polarization-independent, and focus-tunable liquid crystal diffractive lens for augmented reality

Mareddi Bharath Kumar[1], Daekyung Kang[1], Jihoon Jung[2], Hongsik Park[2], Joonku Hahn[2], Muhan Choi[2], Jin-Hyuk Bae[2], Hyunmin Kim[3], and Jonghoo Park[1*]

[1]Department of Electrical Engineering, Kyungpook National University, Daegu, Republic of Korea

[2]School of Electronics Engineering, Kyungpook National University, Daegu, Republic of Korea

[3]Companion Diagnostics & Medical Technology Research Group, DGIST, Daegu 42988, Republic of Korea.

*Correspondence to: jonghoopark@knu.ac.kr

**I. Lens Fabrication**

The lens fabrication started from cleaning an ITO coated PET substrate with acetone, IPA, and DI water in an ultrasonic bath for 5 min each. The sample was cleaned in UV ozone chamber for 10 min. A photoresist, GXR 601, was spin-coated on the sample with 3500 rpm for 30 sec and soft baked at 110°C for 50 sec. The photoresist was exposed to UV wavelength of 350 nm with an energy of 18 mJ under the mask for Fresnel zone pattern electrode in the mask aligner (MIDAS MDA-8000B). The UV exposed area of the photoresist was developed by AZ300 developer for 50 sec. The photoresist pattern was then hard baked at 110°C for 2 min. The ITO films not covered by the photoresist was etched by using LCE 12K diluted by DI water with 1:3 ratio. After rinsing with DI water, a 500 nm thick SU8 2000.5 layer was spin coated on the sample with 4000 rpm for 30 sec. The SU8 was exposed to UV with energy of 26 mJ under the mask for via holes. The exposed area of the SU8 was developed by SU8 developer for 1 min 10 sec. The sample was then exposed to UV with a power of 10 mJ/sec for 5 min without a mask (flood exposure). The sample was rinsed with acetone, IPA, and DI. A negative image reversal photoresist AZ 5214 was spin coated on the sample with 3500 rpm for 30 sec and baked at 110°C for 1 min. The sample was exposed by UV light with energy of 25 mJ under the mask for aluminum bus lines followed by hard baking at 110°C for 1 min 45 sec. The photoresist was developed in AZ300 developer. The aluminum bus lines were formed by depositing 600 nm thick aluminum layer by using thermal evaporation followed by a lift-off process. The 2wt% of polyvinyl alcohol (PVA) was prepared and mixed with diameter of 10 μm bead spacers. The PVA was then spin coated on the sample and baked at 90°C for 1 hr. The PVA was also spin coated on ITO coated PET to fabricate a top substrate. The PVA layers on top and bottom substrates were rubbed by using a velvet cloth. A 10 μm thick adhesive spacers were placed on the rim of the bottom substrate. The top and bottom substrates were then bonded in antiparallel rubbing direction. A commercial E7 liquid crystal was filled into the cell by capillary action at 60°C.

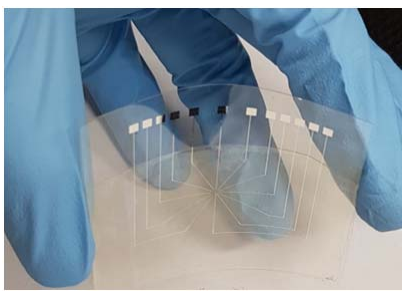

**Figure S1.** Image of the lens

## II. Voltage–dependent optical transmission measurement

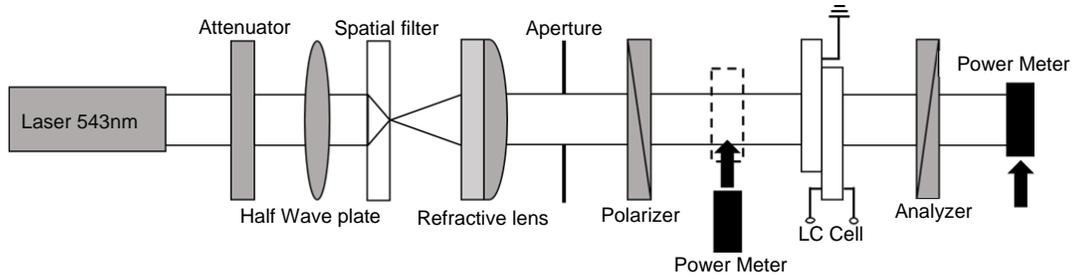

**Figure S2**. Measurement setup for the voltage–dependent optical transmission.

Figure S2 shows the measurement setup for voltage–dependent optical transmission of the homogeneously aligned LC between PET substrates. The setup is comprising a 543 nm wavelength He-Ne laser, an attenuator (ATT), a half wave plate, a spatial filter, a refractive lens (with focal length of 200 mm), an aperture, a polarizer, the LC lens, and an analyser. The intensity of the light incident on the LC was monitored by the optical power meter and kept constant by adjusting an attenuator and a half wave plate regardless of the orientation of optical axes of the polarizer. The spatial filter consists of a microscope objective, a pinhole aperture, and a positioning mechanism. The refractive lens was used to collimate the beam. The angle between the optical axes of the polarizer and the LC was changed by rotating the optical axes of the polarizer and the anlayzer. The angle between the polarizer and the analyser was kept perpendicular to each other during rotation. The optical transmission was measured for the different angles ($\varphi$) between the optical axes of LC and the polarizer as a function of the voltage applied to the LC. All 12 electrodes were electrically connected during the measurement.

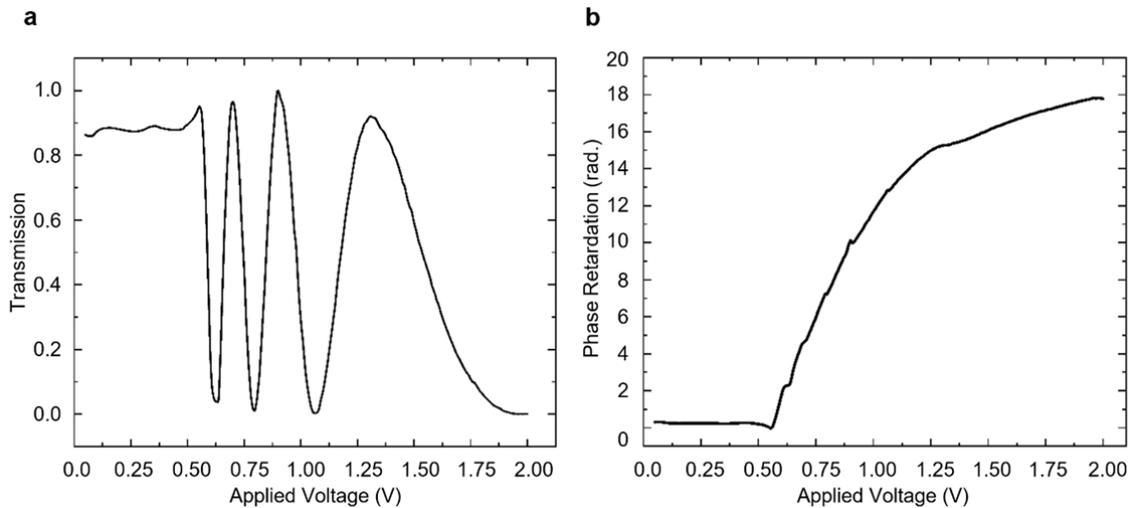

**Figure S3**. Voltage–dependent (a) optical transmission and (b) phase retardtion of PET(0°)/LC(45°)/PET(90°) structure as a function of voltages applied to the LC.

Figure S3a shows the voltage–dependent optical transmission with angle between polarizer and LC lens of 45 degrees. The optical transmission oscillates between the minimum and maximum value as a function of the voltage applied to the LC as the birefringence of the LC goes through full and half wave plate conditions. The voltage–dependent optical transmission can be translated into phase retardation of LC, and the relation between optical transmission in crossed polarizers ($I_{cross}$) and phase retardation ($\delta$) is described by

$$I_{cross}=\sin^2(2\varphi)\sin^2(\delta/2), \qquad (1)$$

Where $\varphi$ is the angle between optical axis of a polarizer and extraordinary axis of LC. Figure S3b shows the phase retardation as a function of voltage extracted from Figure S3a.

### III. Jones Matrix Equation

The retardation in birefringent material can be modelled as a retardation matrix. We consider a PET and liquid crystal as retarders that introduce a phase difference ($\Delta$) between ordinary and extraordinary rays. The retardation could be considered as $-\frac{\Delta}{2}$ along the slow axis and $+\frac{\Delta}{2}$ along the fast axis. Thus, the incident light on a retarder can be represented by following complex number notation;

$$P = \begin{bmatrix} e^{-i\frac{\Delta}{2}} & 0 \\ 0 & e^{+i\frac{\Delta}{2}} \end{bmatrix}$$

A retarder introduces a rotation ($\gamma$) and retardation ($\Delta$), where $\gamma$ is the angle between the ordinary axis of the PET and optical axis of the incident light. The orientation of extraordinary and ordinary axes of the retarder with respect to the optical axis of incident light can be represented as in the figure S4.

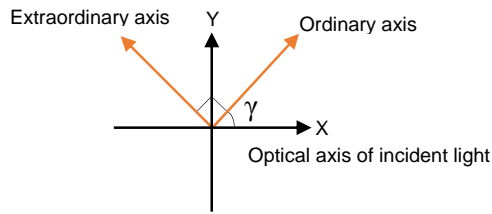

**Figure S4**. The orientation of extraordinary and ordinary axes of the retarder with respect to the reference axis.

Jones matrix for a retarder can be represented by following matrix.

$$T = R(-\gamma) \, P \, R(\gamma) = \begin{bmatrix} \cos(\gamma) & -\sin(\gamma) \\ \sin(\gamma) & \cos(\gamma) \end{bmatrix} \begin{bmatrix} e^{-\frac{\Delta}{2}} & 0 \\ 0 & e^{\frac{\Delta}{2}} \end{bmatrix} \begin{bmatrix} \cos(\gamma) & \sin(\gamma) \\ -\sin(\gamma) & \cos(\gamma) \end{bmatrix} \quad (2)$$

$$T = \begin{bmatrix} \cos\frac{\Delta}{2} - i\sin\frac{\Delta}{2}\cos 2\gamma & -i\sin\frac{\Delta}{2}\sin 2\gamma \\ -i\sin\frac{\Delta}{2}\sin 2\gamma & \cos\frac{\Delta}{2} + i\sin\frac{\Delta}{2}\cos 2\gamma \end{bmatrix} \quad (3)$$

Where T is the transmission of light. $R(\gamma)$ and $R(-\gamma)$ are the rotation and reverse rotation matrix, respectively. The reverse rotation matrix is used to refer back to the original reference axis (X, Y). The optical transmission for the structure of PET(0°)/LC(45°)/PET(90°), $E_{out}$, can be calculated by multiplying Jones matrix for PET1, LC, and PET2.

$$PET1 = \begin{bmatrix} \cos\frac{\Delta_1}{2} - i\sin\frac{\Delta_1}{2}\cos 2\gamma & -i\sin\frac{\Delta_1}{2}\sin 2\gamma \\ -i\sin\frac{\Delta_1}{2}\sin 2\gamma & \cos\frac{\Delta_1}{2} + i\sin\frac{\Delta_1}{2}\cos 2\gamma \end{bmatrix} \quad (4)$$

$$LC = \begin{bmatrix} \cos\frac{\Delta_2(V)}{2} + i\sin\frac{\Delta_2(V)}{2}\sin 2\gamma & -i\sin\frac{\Delta_2(V)}{2}\cos 2\gamma \\ -i\sin\frac{\Delta_2(V)}{2}\cos 2\gamma & \cos\frac{\Delta_2(V)}{2} - i\sin\frac{\Delta_2(V)}{2}\sin 2\gamma \end{bmatrix} \quad (5)$$

$$PET2 = \begin{bmatrix} \cos\frac{\Delta_1}{2} + i\sin\frac{\Delta_1}{2}\cos 2\gamma & i\sin\frac{\Delta_1}{2}\sin 2\gamma \\ i\sin\frac{\Delta_1}{2}\sin 2\gamma & \cos\frac{\Delta_1}{2} - i\sin\frac{\Delta_1}{2}\cos 2\gamma \end{bmatrix} \quad (6)$$

$$E_{out} = \begin{bmatrix} \cos\frac{\Delta_2(V)}{2} + i\sin\frac{\Delta_2(V)}{2}\sin 2\gamma\cos\Delta_1 & 2\sin\frac{\Delta_1}{2}\cos\frac{\Delta_1}{2}\sin\frac{\Delta_2(V)}{2} - i\sin\frac{\Delta_2(V)}{2}\cos 2\gamma\cos\Delta_1 \\ -2\sin\frac{\Delta_1}{2}\cos\frac{\Delta_1}{2}\sin\frac{\Delta_2(V)}{2} - i\sin\frac{\Delta_2(V)}{2}\cos 2\gamma\cos\Delta_1 & \cos\frac{\Delta_2(V)}{2} - i\sin\frac{\Delta_2(V)}{2}\sin 2\gamma\cos\Delta_1 \end{bmatrix} \quad (7)$$

where $i$ is the imaginary unit, $\Delta_1 = 2\pi \cdot \delta n \cdot d_{PET}/\lambda$ is the phase retardation of the PET, $\delta n$ and $d_{PET}$ are the birefringence and thickness of the PET, $\Delta_2 = 2\pi \cdot \delta n(V) \cdot d/\lambda$ is the phase retardation of the LC, $\delta n(V) = n_{eff}(V) - n_0$, $1/n_{eff}^2(V) = \cos^2\theta(V)/n_e^2 + \sin^2\theta(V)/n_o^2$, $\theta(V)$ is a tilt angle between the plane perpendicular to the propagating direction of the ray and the optical axis of the LC, V is the voltage across the LC, d is the thickness of the LC, and $\lambda$ is the wavelength of incident light, $\gamma$ is the angle between ordinary axis of PET1 and polarization of incident light.

For the PET with a thickness of 128.5 μm, the matrix (7) becomes independent of the angle between optical axis of incident light and ordinary axis of the PET as shown below.

$$E_{out} = \begin{bmatrix} \cos\frac{\Delta_2}{2} & \sin\frac{\Delta_2}{2} \\ -\sin\frac{\Delta_2}{2} & \cos\frac{\Delta_2}{2} \end{bmatrix}$$

## IV. RGB imaging using the ultrathin lens with model eye with ±3.0 D

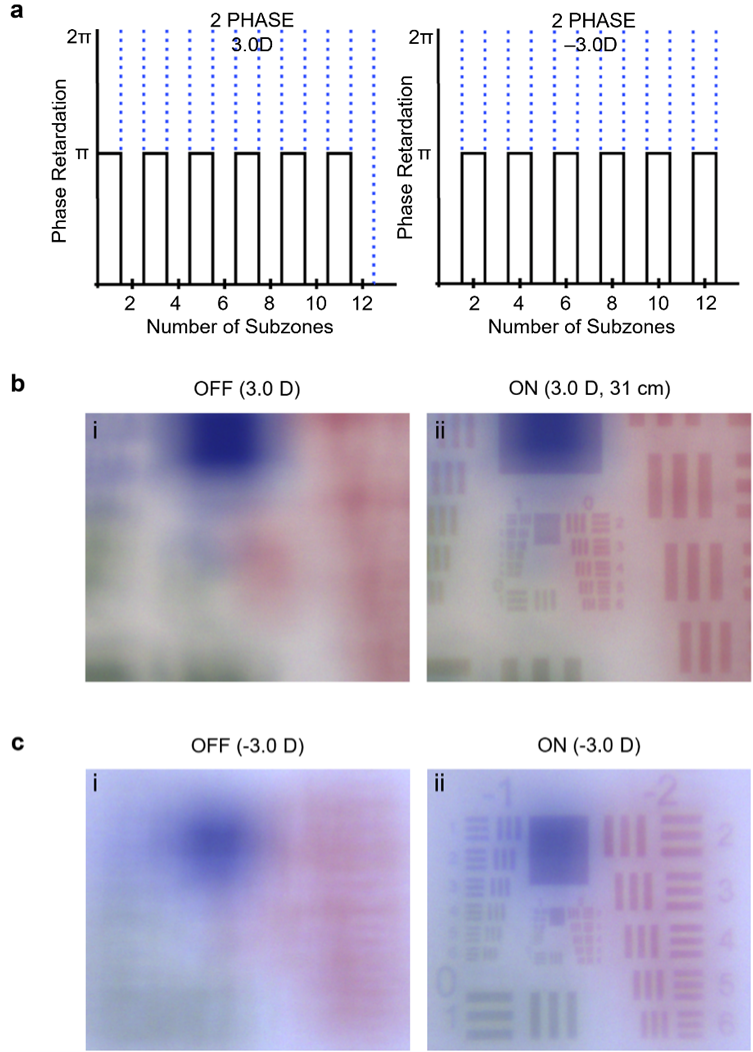

**Figure S5**. RGB imaging using the ultrathin lens (a) Phase profile for the positive and negative 3D power. (b) Images of USAF resolution target placed at 31 cm when the LC lens was turned off (left) and on (right) with optical power of +3.0 D. (c) Images of USAF resolution target placed at 60 cm when the LC lens was turned off (left) and on (right) with optical power of -3.0 D. The meniscus lens with optical power of +3.0 D was placed in front of the LC lens.